\documentstyle [12pt]{article}
\begin{document}
\centerline {CONTRIBUTION TO THE CONFERENCE}
\vskip 1mm
\centerline {"NEUTRINO PHYSICS AND ASTROPHYSICS", Eilat May 29 - June 3 , 1994}
\vskip 2cm
\centerline {\bf LUMINESCENT BOLOMETER AND NEUTRINO PHYSICS}
\vskip 2cm
\centerline {\bf L. GONZALEZ-MESTRES}
\vskip 4mm
\centerline {Laboratoire d'Annecy-le-Vieux de Physique des Particules, }
\centerline {B.P. 1110 , 74941 Annecy-le-Vieux Cedex, France}
\vskip 1mm
\centerline {and}
\vskip 1mm
\centerline {Laboratoire de Physique Corpusculaire, Coll\`ege de France,} 
\centerline {11 place Marcellin-Berthelot, 75231 Paris Cedex 05 , France }
\vskip 3cm
The luminescent bolometer, proposed in 1988 , 
is now seriously considered for several applications
in nuclear and particle physics: dark matter searches, 
double beta decays, low energy neutrino physics,
heavy ion physics... It is also a very promising device for basic 
condensed-matter physics and chemistry
experiments, and may lead to astrophysical applications. 
The luminescent bolometer is based on the simultaneous
detection of light and phonons, allowing for particle 
identification and for a detailed study of the detector response.
Digitized analysis of the signals produced in several sensors installed on the 
same crystal is then a very 
powerful tool.
Superconducting sensors allow to detect the scintillation light pulse followed
by the delayed front of phonons, and can be extremely sensitive 
leading to single photon counting in the visible range.
They also provide information on the position of the event inside the absorber, and can be fast enough for all
proposed applications. The luminescent bolometer, with superconducting sensors, appears extremely promising for real time $solar$
$neutrino$ $experiments$ based on new $indium$ single crystal scintillators.
We focus on this particular application, discussing the status of the art as 
well as open problems and presenting
an updated description of a full scale real time solar neutrino experiment 
sensitive to the low energy sector.
\vskip 3mm
Other applications of the luminescent bolometer (e.g. 
spectroscopy or neutrino detection at reactors), involving indium compounds and
other single crystal scintillators, are equally considered and discussed in 
detail.
\vskip 8cm
{\bf 1. The Luminescent Bolometer}
\vskip 6mm
Simultaneous detection of light and phonons in a single crystal scintillator 
cooled to very low temperature 
was proposed in 1988 $^1$
as a new tool for particle detection. The new technique was expected to: 
a) provide slow thermal detectors with a fast light
strobe, allowing for a much better timing without crucially spoiling energy 
resolution; b) 
make possible particle identification through the phonon/light ratio, 
thus improving background rejection.
As thermal bolometers evolve nowadays towards the detection of nonequilibrium 
phonons, fluorescence appears as a natural
complement. Many crystals are expected to produce an important light yield at 
very low temperature.
The device operating simultaneous detection of light and phonons at very low 
temperature was called the $luminescent$
$bolometer$ $^2$.
Simultaneous detection of fluorescence light and nonequilibrium phonons would 
allow $^{3,4}$ to combine good
energy resolution, fast timing, position information and high background 
rejection, leading to a new generation of very 
performant devices. Several extremely difficult experiments may become 
feasible, whereas a certain number of present day 
experiments would be seriously improved by the new cryogenic device.
\vskip 3mm
A number of well-known scintillators (BGO, CdWO$_4$ , CaWO$_4$ , GSO:Ce, 
CeF$_3$ , YAG:Ce, CaF$_2$:Eu...) exhibit fast luminescence
at low temperature and are serious candidates for absorbers. 
But other substances, which do not scintillate at room
temperature, become efficient luminophores when cooled down
as thermal quenching disappears. Two examples may be particularly 
relevant to particle detection:
\vskip 3mm
a) PbMoO$_4$ green fluorescence is known $^{5,6}$ to increase by four orders of magnitude between room temperature
and LN$_2$ temperature. Its fluorescence properties have been studied down to 
He$_4$ temperature $^6$. Following a proposal
to apply cooled PbMoO$_4$ to double beta experiments $^7$, a PbMoO$_4$ $2 
\times 2 \times 2$ cm$^3$ single crystal
read by a photomultiplier through a quartz light guide was 
characterized down to LN$_2$ temperature and showed a
photopeak pulse height equal to 16$\% $ of that of room temperature 
NaI:Tl $^8$.
\vskip 3mm
b) Some indium oxides studied by J.P. Chaminade $^9$ following a proposal 
$^{10,11,4}$ to develop a scintillating single
crystal incorporating $^{115}$In as a basic element, 
exhibit encouraging scintillating properties at low temperature.
An example is In$_2$Si$_2$O$_7$ $^{10,12}$ which has been characterized down to 4K $^{12}$, and
many other compounds can be considered $^{10}$.
\vskip 3mm
If the properties of the absorber (Debye temperature, phonon propagation, 
low temperature scintillation light yield and decay
time...) are crucial to the quality of a luminescent bolometer, the sensors 
are equally key elements. The firts design
proposed $^{1,2}$ the use of separate sensors for the light and the phonons. 
Semiconductors, 
thin black bolometers and superconductors were considered as photon sensors 
$^{1,2,11}$.
The first successful feasibility study, made by the Milano group, 
adopted such an approach $^{13,14}$ using 
a photodiode as the cryogenic photosensitive device. This practical 
solution allowed
to demonstrate the principle without previously undertaking long technical 
developments.
The results were naturally limited in
threshold and energy resolution, as has been discussed in recent reviews 
$^{15,16}$,
but the situation can be improved. In 1991 , we proposed $^{3,4}$
a new design incorporating: a) a common sensor for both the light strobe and 
the delayed phonon signal;
b) the use of arrays of superconducting
tunnel junctions (STJ) as the new sensor.
\vskip 1cm
{\bf 2. Relevance of superconducting sensors}
\vskip 6mm
Superconductors are natural sensors for both phonons and photons, and should 
perform better than semiconductors due to
the comparatively low gap for quasiparticle excitation.
Low impedance superconducting films already provide the best phonon sensors 
$^{17}$,
and photosensitive superconducting devices are an active research 
subject $^{18}$.
A performant superconducting sensor, sensitive to the light strobe 
followed by the delayed pulse of phonons,
would considerably simplify and improve the architecture of the luminescent 
bolometer.
This seems feasible nowadays due to the success of arrays of superconducting 
tunnel junctions $^{19}$ and of other
superconducting sensors $^{20}$. The new device, 
made of a single crystal low temperature scintillator with an appropriate 
superconducting sensor implanted on each of its faces,
may become the ultimate detector for several physics goals.
\vskip 4mm
$2a.$ $STJ$ $arrays$
\vskip 5mm
With a series array of 432 Al-Al STJ, implanted on a Si wafer with an area of 
12 $\times $ 12 mm$^2$ and a thickness of 0.5 mm,
the Oxford group obtained $^{19}$ at T $\simeq $ 360 mK (base temperature of 
the cryostat) a resolution
of 700 eV FWHM on a 25 keV X-ray peak produced from the fluorescence of an 
indium foil. 
A naive extrapolation suggests $^{3,4}$ that a similar
STJ array could be sensitive to $\approx $ 1 keV of photons absorbed near 
the array, which corresponds to the light
yield of a $\approx $ 10 keV electron or photon in an efficient scintillator.
According to the Oxford group$^{19}$, most of the 25 keV peak width was due 
to drifts in the cryostat temperature 
and to electronic noise.
These considerations motivated our proposal to use arrays of STJ for the 
detection
of both photons (the fast strobe) and phonons (the delayed pulse).
To absorb the scintillation light we considered, either implanting a thin layer between the radiation absorber and the STJ 
array (but care must be taken of phonon propagation through the
layer), or to use blackened STJ arrays covering a large fraction of the crystal surface
(which requires working in an optical cavity).
\vskip 3mm
An interesting possibility would be to deposit the STJ arrays on a 
superconducting substrate
of higher critical temperature, covering the full crystal surface
(an array per face). Nb or Sn can be the substrate for a Al STJ array.
Thus, both the photons and the phonons from the absorber would be 
converted into quasiparticles
by the substrate layer in an efficient way. 
Such quasiparticles would subsequently be detected by the STJ array.
The photon signal would immediately originate in the substrate,
whereas phonons would first undergo a number of scattering processes 
depending on the size and quality of the crystal.
With an ADC and a DSP after the electronic chain, for each face of 
the crystal, digital analysis
would allow to reach a very low threshold for both the fluorescence 
and the phonon signal. It should be noticed that,
when the expected signal is a sum of exponentials, on-line digital 
filtering allows for iterative algorithms
leading to very performant trigger schemes $^{21}$,
which apply to fluorescence in a straightforward way and can be adapted 
to phonon
detection. Energy resolution would also be very good, 
as total energy can be reconstructed from light and phonon pulses. 
Digital analysis
of the phonon pulse would lead to excellent space resolution inside the crystal.
Timing would depend on the fluorescence properties and on the superconducting read-out, but it seems reasonable 
to expect to reach fast timing (down to 100 ns) with suitable choices.
However, the crystal size and phonon scattering properties will necessarily set an intrinsic limit to the detector performance.
Phonons reaching the crystal surface with an energy E $<$ 2$\Delta $ 
(the gap of the STJ 
superconducting material) cannot contribute to the signal.
\vskip 3mm
Even if the scheme and goals are not identical, it is worth noticing that
Perryman et al. propose$^{22}$ optical photon counting with a superconducting 
substrate in combination with an array of
widely spaced STJ of lower energy gap.
In our case, we are not interested in optical photon counting but in the 
detection of 10$^2$ - 10$^5$ optical photons
produced by a particle interacting with the cooled scintillator. 
On the other hand,
we must face the extra requirement of efficient phonon detection from large 
absorbers.
The fact that several intrinsic scintillators work at low temperature 
is encouraging,
as doped scintillators may exhibit poor phonon propagation.
An attempt to model non-equilibrium signals in a series array of STJ has 
recently been performed by the Oxford group
$^{23}$.
\vskip 5mm
$2b.$ $Other$ $techniques$
\vskip 5mm
Superconducting films are very successful $^{17, 24}$ and solutions based on 
this technique, others than STJ arrays,
deserve serious consideration. But efficient detection of scintillation 
light is likely to
limit the freedom of the design.
It is possible to consider superheated superconducting dots$^{25}$ when 
only four faces of the crystal need to be used
(allowing for a magnetic field parallel to the four faces),
or for cylinder structures. But this may limit space resolution.
Sensitivity may also be a problem.
It seems difficult to simultaneously detect light and external phonons using 
superheated microspheres,
but we may hope for technical progress in the interface between the granules 
and the scintillating absorber.
\vskip 5mm
$2c.$ $A$ $critical$ $remark$
\vskip 5mm
Although luminescence properties are not expected to drastically change between 4K and dilution temperatures, 
it must be noticed that only two experiments $^{13,14,26}$ have studied the 
luminescence of a thermal bolometer cooled to very
low temperature, and only one of them $^{13,14}$ incorporated a sensor for the 
scintillation light. Many more
experiments are required on luminescence at very low temperature before setting the design of a superconducting sensor for
light and phonons. However, STJ arrays present the advantage of operating with 
excellent
performances at He$_3$ and even at He$_4$ temperatures, which simplifies 
several basic problems.
Simultaneous detection of light and phonons presents a similar advantage as 
compared to simultaneous detection of ionization
and heat.
\vskip 1cm
{\bf 3. Proposed applications}
\vskip 6mm
$3a.$ $Non$-$baryonic$ $dark$ $matter$
\vskip 5mm
This was the first proposed application of the luminescent bolometer $^2$, in 
view of background rejection and nucleus recoil
identification. The approach has recently been criticized on the grounds of the high threshold of existing prototypes
$^{16}$, but as has been explained in this paper the present situation can be 
considerably improved introducing superconducting
sensors: then the threshold of the luminescent bolometer will become as low as 
that of any device performing 
simultaneous detection of ionization and heat. The possibility to work well 
above dilution temperatures will then become
a definite advantage of dark matter experiments using the luminescent 
bolometer.
The reliability of a large scale experiment would be much better with our 
approach, where 100 kg to 1 ton detectors can indeed
be cooled to the operating temperature with existing and well established 
cryogenic techniques. Furthermore,
targets such as $^7$Li, $^{19}$F, $^{27}$Al, $^{127}$I, $^{183}$W... can be 
incorporated in the cold scintillator approach.
If particle physics and cosmology still provide a ground to experiments aiming 
at the direct detection of dark matter WIMPs
(there is to date no evidence for new particles!),
the luminescent bolometer with a superconducting read-out is to be the right 
technique for that purpose.
However, many basic studies remain to be performed on the low temperature 
behaviour of the relevant scintillators.
\vskip 5mm
$3b.$ $Double$ $beta$
\vskip 5mm
Applications to double beta experiments were proposed in 1989 $^7$, with the 
idea to reject the alpha background in high 
Q materials. CdWO$_4$ and PbMoO$_4$ were then explicitly considered.
The use of CdWO$_4$ seems indeed to be a promising way $^{27, 15}$. To our 
original proposal, the Milano group has added
the successful development of a CaF$_2$ luminescent bolometer $^{13,14}$.
The use of superconducting sensors would be crucial to improve energy 
resolution, allowing for: a) a better background rejection
(rejection of $\alpha $'s, but also eventually separation between $\beta $ or 
$\gamma $ and $2\beta $ events);
b) eventually, the necessary separation 
in energy between the tail of double beta decays with neutrinos and 
neutrinoless double beta 
events, feasible only through energy resolution and possibly through a careful 
analysis of the phonon pulse sensitive to the details
of electron energy losses.
Comparison between the fluorescence and the phonon pulse will be crucial, not 
only because of phonon/light ratio but also
through the study of the delay between both pulses on the six faces of the 
crystal.
With these tools, a molybdenum experiment using a molybdate cooled to very low 
temperature should be seriously considered.
After suitable technical developements, the luminescent bolometer can 
potentially incorporate any double beta target.
The elementary cells 
of a double $\beta $ experiment 
can be $\approx $ 2 $\times $ 2 $\times $ 2 cm$^3$ crystals, which amounts to 
$\approx $ 25 crystals and
$\approx $ 150 electronic channels per Kg of detector.
\vskip 5mm
$3c.$ $Solar$ $and$ $reactor$ $neutrinos$ $(indium$ $target)$
\vskip 5mm
This may become the most important and far-reaching application of the 
luminescent bolometer, as no technique allows
by now to detect in real time low energy neutrinos.
Recent progress on indium single crystal scintillators is encouraging $^9$, but it is unlikely that
a room temperature scintillator would lead to a correct background 
rejection$^{3,4}$.        
2$\beta $ and 3$\beta $ coincidences from $^{115}$In radioactivity, as 
well as coincidences between a $^{115}$In $\beta $ and
an erratic gamma, are the main worries for a solar neutrino detector with 
a $^{115}$In target.
Room temperature scintillation does not seem to provide enough information to 
fight such a background, but detecting phonons
in addition would considerably improve space and energy resolution, which are 
crucial to evaluate the detector performance.
\vskip 4mm
Several indium compounds
seem to scintillate mainly at low temperature. 
To the indium germanates, silicates and other oxides presently under study,
some of which (e.g. In$_2$Si$_2$O$_7$) give excellent results at 
low temperature, InCe oxides should be added in order to possibly
exploit the fluorescence of trivalent cerium. Fluorides deserve further 
consideration$^{9,10}$, as some of them can scintillate
and crystal growth seems easier than with oxides.
A large scale, real time solar neutrino experiment
based on Raghavan'as reaction, with a luminescent bolometer made of an indium
compound, has already been described $^{3,4,28}$ and nothing to date 
contradicts its potential feasibility.
With $\approx $ 1cm space resolution
in 4 $\times $ 4 $\times $ 4 cm$^3$ single crystals, segmentation in a 
few million elementary cells would be achieved, allowing for a
signature based on the triple coincidence between the $^{115}$ $\beta $ and the two, spatially separated, delayed $\gamma $'s
(116 and 496 keV).
But the large number of crystals ($\sim $ 3.10$^4$) and electronic channels 
($\sim $ 2.10$^5$) required, with on line digital
analysis, makes it a great challenge even if the appropriate scintillating 
crystals become available.
\vskip 4mm
A neutrino-antineutrino oscillation experiment at a reactor would be $\sim $ 
100 times smaller.
It must therefore be considered as a preliminary step, working at a higher 
neutrino energy (higher inverse $\beta $ energy)
which should considerably simplify the above discussed background rejection. 
On the other hand, ambient background is
much worse and deserves a careful study where self-sheilding (active) is likely to play a crucial role.
\vskip 4mm
The possibility to actually find
performant intrinsic indium single crystal scintillators is real and 
exciting $^9$. 
Such a breakthrough would considerably improve the
prospects for phonon detection in large scintillating indium crystals. 
However, light yield and fluorescence lifetime
under $\beta $ and $\gamma $ irradiation remain a challenge,
as very sharp performances are required in the proposed experiments.
Superconducting sensors appear as the only solution to build a feasible 
detector, satisfying all requirements in view 
of background rejection.
\vskip 5mm
$3d.$ $Other$ $applications$
\vskip 5mm
Basic physics and chemistry, as well as nuclear physics and technology, 
are generating many potential applications
of the luminescent bolometer. The study of relaxation phenomena would 
considerably benefit from such a progress
in instrumentation, allowing to detect in real time several components of the 
degraded energy.
Nuclear spectroscopy requiring particle
identification (phonon analysis may even allow to distinguish between 
a $\beta $ and a $\gamma $ inside a crystal),
neutron detection at low counting rate (e.g. with a 
lithium target$^{3,26}$), low radioactivity measurements including study 
and detection of long lived isotopes... are applications of
increasing interest in both scientific and industrial domains, where detector 
performance can be combined with self-shielding$^3$
potentialities. 
Heavy ion physics also appears as an ideal domain for the new device
$^{28}$, as the luminescent bolometer may replace with success all kinds of 
detectors of present-day experiments
(semiconductors, scintillators, forward detectors...).
The luminescent bolometer may become relevant to nuclear and particle physics 
and technology, astrophysics and space
science and technology, material science, environment, biology and medical uses.
Superconducting sensors as well as on line digital analysis are essential tools to fully exploit the potentialities of a
detector whose basic performance will be the ability to record 
in real time a large amount of physical information on
the interaction of particles with matter.
\vskip 1cm
{\bf 4. References}
\vskip 6mm
1. L. Gonzalez-Mestres and D. Perret-Gallix, in "Low Temperature Detectors for 
Neutrinos and Dark Matter - II", Proceedings
of LTD-2 Annecy May 1988 , Editions Fronti\`eres.
\par
2. L. Gonzalez-Mestres and D. Perret-Gallix, Proceedings of the XXIV 
International Conference on High Energy Physics,
Munich August 1988 , Ed. Springer-Verlag, p. 1223 .
\par
3. L. Gonzalez-Mestres, in "Low Temperature Detectors 
for Neutrinos and Dark Matter - IV", Proceedings of LTD-4, 
Oxford September 1991 , Ed. Fronti\`eres, p. 471-479 .
\par
4. L. Gonzalez-Mestres, Proceedings of TAUP 91 , Toledo September 1991 , 
Nuclear Physics B (Proc. Suppl.) 28A (1992), p. 478-481 .
\par
5. Hj. Bernhardt, Phys. Stat. Sol. (a) 91 (1985), p. 643 .
\par 
6. W. Van Loo and D.J. Wolterink, Phys. Lett. 47A (1974), p. 83 .
\par
7. L. Gonzalez-Mestres and D. Perret-Gallix, Moriond Workshop "The Quest for 
Fundamental Constants in Cosmology", March 1989 ,
Ed. Fronti\`eres, p. 352-354 .
\par
8. M. Minowa, K. Itakura, S. Moriyama and W. Ootani, University of Tokyo 
preprint UT-HE-92/06 (1992).
\par
9. See, for instance, T. Gaewdang, Thesis Universit\'e de Bordeaux 
I "Cristallochimie et luminescence de quelques oxides et 
fluorures d'indium", November 1993 .
\par 
10. L. Gonzalez-Mestres and D. Perret-Gallix, Proceedings of the "Rencontre sur la Masse Cach\'ee", Annecy July 1987 ,
Ed. Annales de Physique, p. 181-190 .
\par
11. L. Gonzalez-Mestres and D. Perret-Gallix, Nucl. Instr. and Meth. 
A279 (1989), p. 382 .
\par
12. T. Gaewdang et al. , to appear in Z. Anorg. Allg. Chem., 
and in reference 9 .
\par
13. A. Alessandrello et al., Proceedings of LTD-4 , p. 367 .
\par
14. A. Alessandrello et al., Proceedings of the Moriond Workshop "Progress 
in Atomic Physics, Neutrinos and Gravitation" January 
1992 , Ed. Fronti\`eres, p. 201 .
\par
15. E. Fiorini, Proceedings of LTD-5 , Berkeley July-August 1993 , Journal of 
Low Temperature Physics 93 (1993), 
Numbers 3/4 , p.189 .
\par
16. B. Sadoulet, same Proceedings, p. 821 .
\par
17. See, e.g. P. Ferger et al. "A Massive Cryogenic Particle Detector with Good Energy Resolution", Max-Planck-Institut
preprint Munich 1994 .
\par
18. A. Barone and M. Russo in Advances in Superconductivity, Plenum 1993 .
\par
19. See, e.g. D.J. Goldie, Proceedings of the Workshop on Tunnel Junction 
Detectors for X-rays, Naples December 1990 ,
World Scientific.
\par
20. See, e.g. Proceedings of LTD-5 .
\par
21. L. Gonzalez-Mestres, unpublished; see S. Dil, Rapport de stage 
DESS Universit\'e Jean Monnet, Saint-Etienne June 1992 .
\par
22. M.A.C. Perryman, C.L. Foden and A. Peacock, ESA preprint September 1992 .
\par
23. A.D. Hahn et al. in LTD-5 , p. 611 , and R.J. Gaitskell et al., in LTD-5 ,
p. 683 .
\par
24. V. Nagel et al. in LTD-5 , p. 543 , and H. Kraus et al., in LTD-5 , p. 533 .
\par
25. C. Berger et al. , same Proceedings, p. 509 .
\par
26. P. de Marcillac et al., Nuclear Instruments and Methods A 
337 (1993), p. 95 .
\par
27. Y. Zdesenko et al., Proceedings of the Moriond Workshop "Progress in 
Atomic Physics, Neutrinos and Gravitation"
January 1992 , Editions Fronti\`eres, p. 183 .
\par
28. L. Gonzalez-Mestres, same Proceedings, p. 113-118 .
\end{document}